\documentclass[prb,preprint]{revtex4-2}

\usepackage{ascmac}
\usepackage[dvipdfmx]{graphicx}
\usepackage{dcolumn}
\usepackage{bm}
\usepackage{color}
\usepackage{here}
\usepackage{braket}
\usepackage{amsmath,amssymb}

\begin{document}

\newcommand{\ham}{\hat{\mathcal{H}}}
\newcommand{\hamd}{\hat{\mathcal{H}}^\prime}
\newcommand{\hamdd}{\hat{\mathcal{H}}^{\prime\prime}}
\newcommand{\kp}{{\bm k \cdot \bm p}}
\newcommand{\Ek}{\mathcal E - \bm k}
\newcommand{\SIAk}{\mathrm{SIA}{\bm k}}
\newcommand{\cSOI}{\mathrm{cSOI}}
\newcommand{\SIAp}{\mathrm{SIA}{\bm p}}
\newcommand{\SIApp}{\mathrm{SIA}{\bm p}^2}
\newcommand{\BIAk}{\mathrm{BIA}{\bm k}}
\newcommand{\BIAp}{\mathrm{BIA}{\bm p}}
\newcommand{\cSOIk}{{\mathrm {cSOI}{\bm k}}}
\newcommand{\sigex}{\mathrm{Si}_x\mathrm{Ge}_{1-x}}
\newcommand{\sige}{\mathrm{Si}_{0.5}\mathrm{Ge}_{0.5}}
\newcommand{\tB}{\theta_{\mathrm B}}
\newcommand{\phiB}{\phi_{\mathrm B}}
\newcommand{\phiD}{\phi_{\mathrm D}}
\newcommand{\dphiD}{\delta\phi_{\mathrm{D}}}
\newcommand{\lambdaB}{\lambda_\mathrm B}

\newcommand*{\bfup}{u_{\bm{k}}^{\mathrm{HH}+}}
\newcommand*{\bfum}{u_{\bm{k}}^{\mathrm{HH}-}}
\newcommand*{\bfupm}{u_{\bm{k}}^{\mathrm{HH}}}
\newcommand*{\vecs}{\braket{\bm{s}_{\bm{k}}^{\mathrm{HH}}}}
\newcommand*{\sgn}{\mathrm{sgn}}
\newcommand*{\ofbloch}[1]{{#1}_{\mathrm{B}}}
\newcommand*{\thb}{\ofbloch{\theta}}
\newcommand*{\phb}{\ofbloch{\phi}}
\newcommand*{\phbtld}{\ofbloch{\tilde{\phi}}}
\newcommand*{\Nphb}{N_{\phb}}
\newcommand*{\xb}{\ofbloch{x}}
\newcommand*{\yb}{\ofbloch{y}}
\newcommand*{\zb}{\ofbloch{z}}
\newcommand*{\vecrb}{\ofbloch{\bm{r}}}
\newcommand*{\mat}[1]{\mathbb{#1}}
\newcommand*{\matrot}{\mat{R}}
\newcommand*{\MAT}[1]{
  \begin{pmatrix}
    #1
  \end{pmatrix}
}
\newcommand*{\RMX}[1]{
  \MAT{
    1 & 0 & 0 \\
    0 & \cos{#1} & -\sin{#1} \\
    0 & \sin{#1} & \cos{#1}
  }
}
\newcommand*{\RMZ}[1]{
  \MAT{
    \cos{#1} & -\sin{#1} & 0\\
    \sin{#1} & \cos{#1} & 0 \\
    0 & 0 & 1
  }
}
\newcommand*{\BV}[2]{
  \MAT{
    \sin{#1}\cos{#2}\\
    \sin{#1}\sin{#2}\\
    \cos{#1}
  }
}

\newlength{\wfig}
\setlength{\wfig}{16.46cm}

\title{Bloch sphere representation for Rabi oscillation driven by Rashba field in the two-dimensional harmonic confinement}

\author{Kaichi Arai}
\author{Tatsuki Tojo}\email{tojo@qms.cache.waseda.ac.jp}
\author{Kyozaburo Takeda}\email{takeda@waseda.jp}
\affiliation{Faculty of Science and Engineering, Waseda University, Shinjuku, Tokyo 169-8555, Japan}
\date{June 17, 2024}

\begin{abstract}
We studied the dynamical properties of Rabi oscillations driven by an alternating Rashba field applied to a two-dimensional (2D) harmonic confinement system. 
We solve the time-dependent (TD) Schr\"{o}dinger equation numerically and rewrite the resulting TD wavefunction onto the Bloch sphere (BS) using two BS parameters of the zenith ($\theta_B$) and azimuthal ($\phi_B$) angles, extracting the phase information $\phi_B$ as well as the mixing ratio $\theta_B$ between the two BS-pole states.

We employed a two-state rotating wave (TSRW) approach and studied the fundamental features of $\theta_B$ and $\phi_B$ over time. The TSRW approach reveals a triangular wave formation in $\tB$. Moreover, at each apex of the triangular wave, the TD wavefunction passes through the BS pole, and the state is completely replaced by the opposite spin state. 
The TSRW approach also elucidates a linear change in $\phi_B$. The slope of $\phi_B$ vs. time is equal to the difference between the dynamical terms, leading to a confinement potential in the harmonic system. The TSRW approach further demonstrates a jump in the phase difference by $\pi$ when the wavefunction passes through the BS pole.

The alternating Rashba field causes multiple successive Rabi transitions in the 2D harmonic system. 
We then introduce the effective BS (EBS) and transform these complicated transitions into an equivalent ``single" Rabi one.
Consequently, the EBS parameters $\tB^{\mathrm{eff}}$ and $\phi_B^{\mathrm{eff}}$ 
exhibit mixing and phase difference between two spin states $\alpha$ and $\beta$, leading to a deep understanding of the TD features of multi-Rabi oscillations. 
Furthermore, the combination of the BS representation with the TSRW approach successfully reveals the dynamical properties of the Rabi oscillation, even beyond the TSRW approximation. 
\end{abstract}

\maketitle

\section{Introduction}

{\it Spin} and {\it charge} form a pair of fundamental physical quanta possessed by an electron.
Conventionally, the charge is controlled by an electric field, whereas spin is controlled by a magnetic field.
However, spin couples naturally with the orbital momentum of electrons owing to the 
``angular momentum."
Focusing on this feature, Rashba \cite{Rashba} theoretically predicted that electron spin could be coupled with an electric field via the spin-orbit interaction (SOI) over half a century ago. 
Nevertheless, it was near the end of the last century that the nano-fabrication technique realized ground-breaking experiments that demonstrated the accuracy of Rashba's theoretical prediction \cite{nitta2, pioro}. 
After the discovery of the Rashba SOI, remarkable progress has been realized in spin and SOI related phenomena such as spin field-effect transistor \cite{das},  the Aharonov-Casher effect with respect to Rashba SOI splitting strength \cite{konig,nitta2}, a single electron coherent oscillations \cite{koppens}, the Rashba SOI coupling in 2D magnetoexcitons \cite{Hakioglu}, a spin filter and analyzer \cite{Aharony}, and spin current and related transport phenomena \cite{spincur1, spincur2, spincur3, spincur4, spincur5, spincur6}.

Spin manipulation by an electric field in quantum dot (QD) systems \cite{revQD1} has attracted significant research attention, because the spin states $\alpha$ and $\beta$ in the QD system are considered promising for the physical realization of artificially controlled quantum bits (qubits) \cite{qcqi0,qcqi1,qcqi2}.
Walls \cite{walls} studied the spin dynamics under the parametric modulation in the presence of spin-orbit coupling.
Nowack et al. \cite{koppens2} have succeeded in the coherent control of the single electron spin in a QD system. 
Echeverr\'{i}a-Arrondo and Sherman \cite{AS} studied the dynamics in a harmonic potential subject to ultra-strong SOI and an external magnetic field. 
Van den Berg et al. \cite{Berg} also observed Rabi oscillations in an electronically controlled spin-orbit qubit in an InSb nanowire. 
However, only a few studies have been conducted on the wavefunction (quantum) phase, particularly in time-dependent (TD) systems. 
Moreover, recent studies have revealed an important relationship between the quantum phases and system topology.

Here, we study the Rabi oscillation driven by an alternating Rashba field confined in the 2D harmonic potential QD, focusing on the TD phase of the wavefunction.
To this end, we reformulated the TD wavefunction onto a Bloch sphere (BS) by employing two BS parameters of the zenith $\tB$ and azimuthal $\phiB$ angles.
The conventional approach projected onto unperturbed eigenstates is useful for understanding the details of the transition processes; however, this approach itself removes phase information.
In contrast, the azimuthal angle $\phiB$ in the BS representation extracts the phase difference between the BS's two-pole (spin) states. Furthermore, the zenith angle $\tB$ indicates their mixing ratio.
As such, the combination of the BS representation with the conventional projection approach is expected to reveal fundamental TD phenomena concerning the mixing and phase of the wavefunction.

In Sec. \ref{eigenstates}, we briefly summarize the electronic states modulated by the static Rashba SOI. 
We then study the spin dynamics driven by the alternating Rashba SOI in Sec. \ref{secdyna}. We show the resulting snapshots (Subsec. \ref{STbunkai}) and discuss the multiple and sequential transitions (Subsec. \ref{statepro}). 
In Sec. \ref{BSrep}, we rewrite the TD wavefunction by employing the BS representation and extract the phase terms in the TD wavefunction. 
To deepen our understanding of the TD features of the BS parameters $\tB$ and $\phiB$, we employ the two-state rotating wave (TSRW) approach.
The combination of the BS representation with the TSRW approach successfully reveals the dynamical properties of the Rabi oscillation, even beyond the TSRW approximation.

\section{Eigenstates}
\label{eigenstates}

\subsection{Harmonic confinement}

We confine an electron by a 2D isotropic harmonic potential of strength $\omega_0$. 
To realize the {\it finiteness} of a practical 2D QD system, we encircle the harmonic potential by the cylindrical hard wall $V_{\rm cyld}$, introducing to the hybrid potential $V_{\mathrm {hyb}}(x, y)$. 
By employing the effective mass ($m^\ast$) approach, the static and unperturbed Hamiltonian ${\hat{\mathcal H}}_0 (x, y)$ is given by 
\begin{align}
\label{H0}
{\hat{\mathcal H}}_0 (x, y)=\frac{{\hat {\bf p}}^2}{2m^\ast}+\frac{1}{2} m^\ast \omega_0^2 (\hat x^2+ \hat y^2) + V_{\rm cyld}\equiv\frac{{\hat {\bf p}}^2}{2m^\ast}+V_{\mathrm {hyb}}(x, y).
\end{align}

Owing to the central force field of the confinement potential $V_{\mathrm {hyb}}(x, y)$, the eigenstates of the unperturbed Hamiltonian in Eq. \eqref{H0} are identified by the radial $n$ and angular $l$ quantum numbers, leading to the Schr\"{o}dinger equation having an eigenstate $\ket{nl}$;
\begin{eqnarray}
\label{TIDSHeq}
{\hat{\mathcal H}}_0 (x, y)\ket{nl}=E_{nl}^{(0)}\ket{nl}.
\end{eqnarray}
We solved the static Schr\"{o}dinger equation \eqref{TIDSHeq} numerically using a finite difference technique. We summarize the calculational details in the present numerical diagonalization in Appendix \ref{cal}.

\begin{figure}[hbtp]
\begin{center}
\includegraphics[scale=0.2]{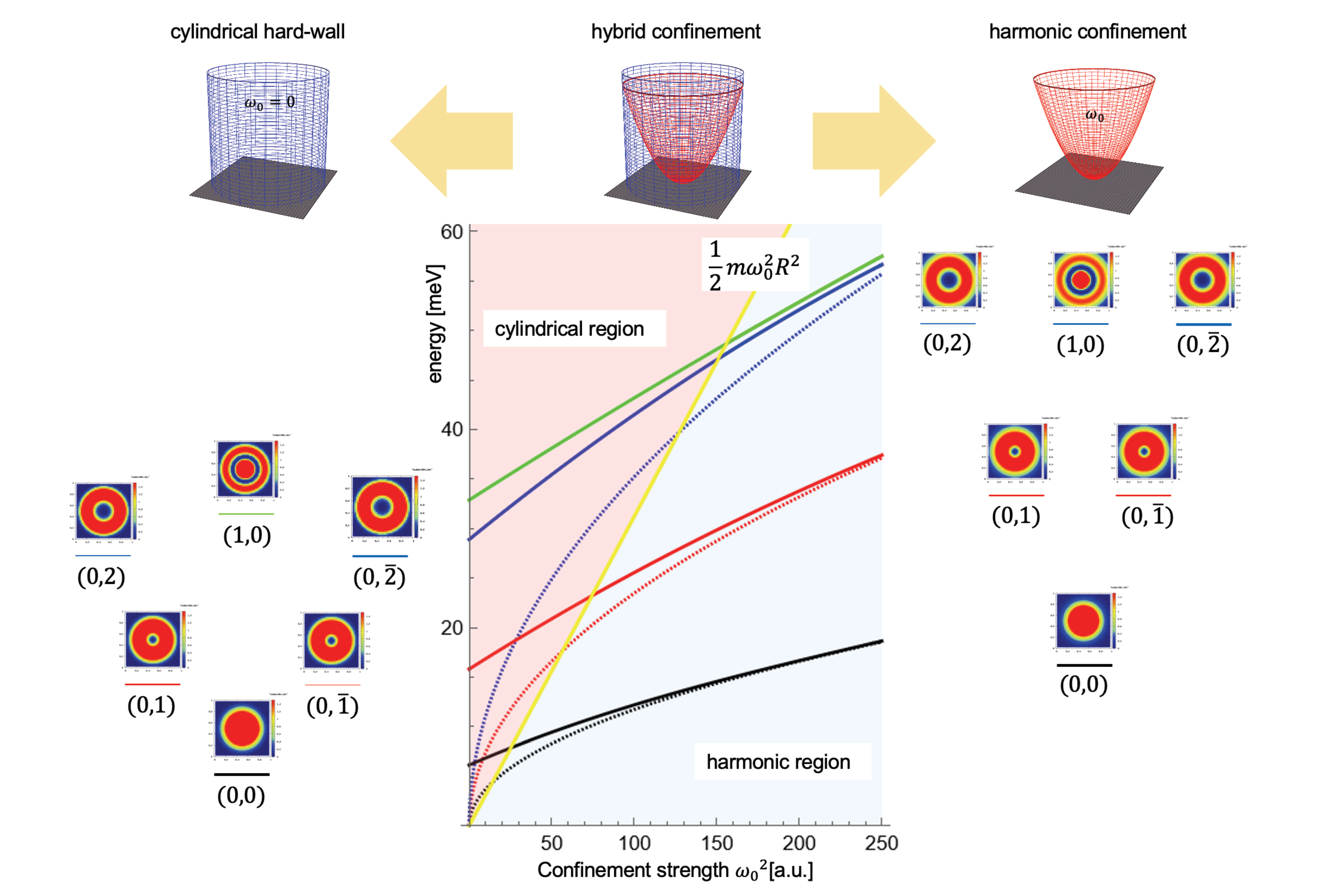}
\caption{Potential profile of the hybrid confinement against the harmonic strength $\omega_0$ reproduced from our previous work \cite{jjap1,jjap2, jjap3, inui} (Copyright 2014, 2016, 2017, The Japan Society of Applied Physics, also 2018, IOP Publishing Ltd); a harmonic potential is surrounded by a cylindrical hard-wall. Solid lines are the energy eigenvalues calculated numerically by the finite difference method and dotted lines indicate those of the ideal harmonic confinement. We also show the resulting electron density profiles of the cylindrical hard-wall ($\omega_0=0$) and the harmonic potential ($\omega_0=15$) in the figure.}
\label{fig1}
\end{center}
\end{figure}

Figure \ref{fig1} shows the calculated eigenstates for the unperturbed Hamiltonian $\hat{{\mathcal H}}_0$ of an InSb 2D QD fabricated on a 100$\times$100 nm square substrate with radius $r_0=50$ nm \cite{Winkler2000}.
The solid lines indicate the eigenstates for the present 2D hybrid QD, whereas the broken lines are the analytical eigenstates in ideal harmonic confinement.
The yellow line indicates the intercept potential between the harmonic and cylindrical confinements. According to Fig. \ref{fig1}, the eigenstates of the ground and lower (first and/or second) excited states is appropriately described by those of the harmonic potential when $\omega_0$ is larger than 15 a.u.

\subsection{Rashba SOI}

The application of a Rashba external electric field ${\boldsymbol \Xi}=(\Xi_x, \Xi_y, \Xi_z)$ to a 2D QD system induces structure-inversion asymmetry and Rashba SOI.
By employing the spin Pauli operator $\hat {\boldsymbol \sigma}$, the Rashba SOI Hamiltonian ${\hat{\mathcal H}}_{\rm R}$ is given by:
\begin{align}
\label{rash}
{\hat{\mathcal H}}_{\rm R}
={\hat {\boldsymbol \sigma}}\cdot ({\boldsymbol \Xi}\times {\hat {\bf p}})
\equiv  {\hat{\mathcal H}}_{\rm R}^+ + {\hat{\mathcal H}}_{\rm R}^- + {\hat{\mathcal H}}_{\rm R}^z, 
\end{align}
with
\begin{align}
\label{HRbunkai}
{\hat{\mathcal H}}_{\rm R}^+ &=- {\Xi_z}(\hat{p}_y +i \hat{p}_x)\hat{\sigma}_+, \nonumber \\
{\hat{\mathcal H}}_{\rm R}^- &=-{\Xi_z}(\hat{p}_y -i \hat{p}_x)\hat{\sigma}_-, \nonumber \\
{\hat{\mathcal H}}_{\rm R}^z &=({\Xi_x}\hat{p}_y - {\Xi_y}\hat{p}_x)\hat{\sigma}_z.
\end{align}
Here, we define the spin step-up ($+$) and step-down ($-$) operators $\hat{\sigma}_\pm$ as: 
\begin{eqnarray}
\hat{\sigma}_\pm \equiv \frac{1}{2}\left( \hat{\sigma}_x \pm i \hat{\sigma}_y \right).
\end{eqnarray}

\begin{figure}[hbtp]\begin{center}
\includegraphics[scale=0.7]{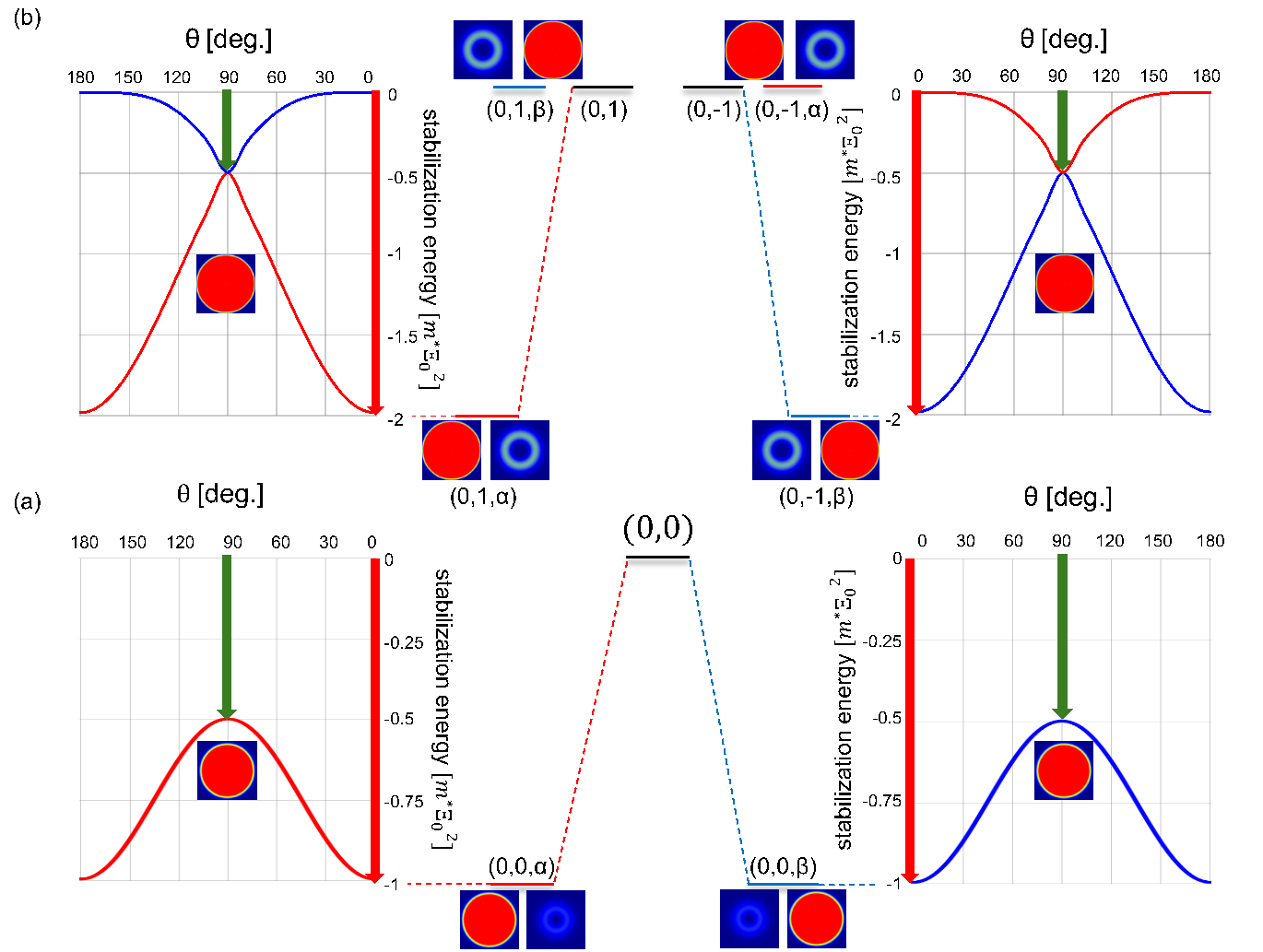}
\caption{Rashba SOI energetics of the ground (a) and 1st excited state (b) reproduced from our previous work \cite{jjap1,jjap2, jjap3,inui} (Copyright 2014, 2016, 2017, The Japan Society of Applied Physics, 2018, IOP Publishing Ltd). We show the energy difference from the unperturbed ground $(n, l)=(0, 0)$ and 1st excited $(0, |1|)$ state over varied $\theta$. The energy difference is normalized by $m^\ast\Xi_0^2$. We also show the spin densities $\rho^\alpha$ (left) and $\rho^\beta$ (right) for the Rashba SOI perturbed state at $\theta =0$. Hereafter, we expediently use the unperturbed notation $(n, l, \sigma)$ for the classification of the perturbed eigenstates. In the present InSb QD, the confinement strength of $1\hbar\omega$ a.u. corresponds to 11.9 meV.}
\label{fig2}
\end{center}
\end{figure}

Figure \ref{fig2} shows the resolution of the spin degeneracy of the unperturbed ground $\ket{00}$ (a) first excited $\ket{01}$ and $\ket{0\bar{1}}$ states (b), in accordance with the change in the Rashba field direction (zenith angle $\theta$). 
The Rashba SOI stabilizes the unperturbed state by mixing a spin component opposite to the original spin. As such, the resolved states have spin components $\alpha$ and $\beta$; for example, the unperturbed ground state $\ket{00}$ with spin $\alpha$ changes into an eigenstate with partly the opposite spin of the $\beta$ component. 
Because the present central force field generates an angular momentum quantum number $l$, the resulting eigenstates should be expressed by the total angular momentum number coupled by $\bm l$ and $\bm \sigma$. 
Nevertheless, we express the eigenstates modulated by the Rashba SOI expediently by $(n l \sigma)$, leading us to employ the spinor representation in later.
Figure \ref{fig2} also demonstrates that the Rashba SOI causes the largest stabilization when the Rashba electric field is applied perpendicular to the 2D plane ($\theta=0$), and the smallest stabilization when $\theta=\pi/2$. 
Accordingly, we studied the case of a perpendicular application as follows. 
It should be noted that the stabilization energies of the Rashba SOI are sufficiently small compared to the unperturbed energies, leading to the appropriateness of the perturbation approach. 
Furthermore, the so-called $\bm l \cdot \bm s$ SOI coupling caused by the central force field potential is negligible because of its smaller size compared to that of the Rashba perturbation by $\sim 10^{-5}$.

\section{Rabi oscillation}
\label{secdyna}

\subsection{TD Rashba field and Snapshots}
\label{STbunkai}

We now vacillate the above Rashba electric field $\hat{\mathcal H}_R(t)$ periodically against time at frequency $w$. 
\begin{eqnarray}
\label{tdHR}
\hat{\mathcal H}_R(t) = \hat {\mathcal H}_{\rm R} \cdot \sin  w t.
\end{eqnarray}
Consequently, the dynamical properties of the confined electron were studied by solving the following TD Schr\"{o}dinger equation:
\begin{align}
\label{td1}
i\hbar\frac{d}{dt} \ket{\psi (t)}=\hat{\mathcal H}(\bm r; t) \ket{\psi (t)}
= \left(\hat{\mathcal H}_0(\bm r) + \hat{\mathcal H}_R(t) \right) \ket{\psi (t)}.
\end{align}
We solve the TD Schr\"{o}dinger equation \eqref{td1} numerically by employing the real-space and real-time grid approach via a 4th-order Runge-Kutta method (Appendix \ref{4RK}).

An application of the Rashba field perpendicular to the 2D plane removes the component ${\hat{\mathcal H}}_{\rm R}^z$ in Eq. \eqref{HRbunkai}, and the perturbed Hamiltonian of $\hat {\mathcal H}_{\rm R}= -\Xi_0 (\hat p_y \hat \sigma_x - \hat p_x \hat \sigma_y)$ results.
For convenience, in the following discussion, we rewrite the time-oscillating Rashba Hamiltonian $\hat{\mathcal H}_R(t)$ as
\begin{eqnarray}
\label{tdHR2}
\hat{\mathcal H}_R(t) = \hat {\mathcal H}_{\rm R} \cdot \sin  w t= \frac{\hat {\mathcal H}_{\rm R}}{-2i}e^{-i wt}
+\frac{\hat {\mathcal H}_{\rm R}}{2i}e^{i wt} \equiv \hat F e^{-i wt} + \hat F ^\dagger e^{i wt},
\end{eqnarray}
where we define the time-independent operator $\hat F$ as: 
\begin{eqnarray}
\label{Fdef}
\hat F = -\frac{\hat {\mathcal H}_{\rm R}}{2i}= \frac{\Xi_0}{2i}(\hat p_y \hat \sigma_x - \hat p_x \hat \sigma_y).
\end{eqnarray}

\begin{figure}[hbtp]
\begin{center}
\includegraphics[scale=0.38]{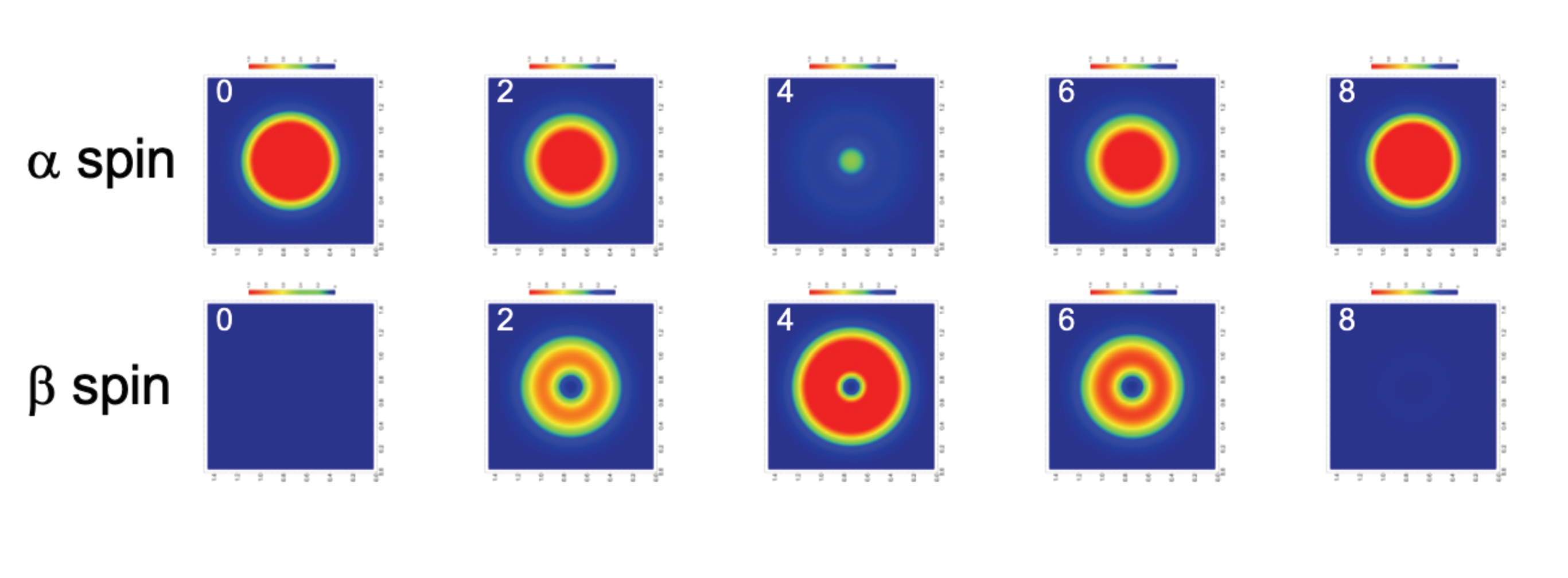}
\caption{Snapshots of TD spin density profiles of $\alpha$ and $\beta$. An electron is initially confined by the harmonic potential $\omega_0=15$ having the spin $\alpha$, as described in text.}
\label{snap}
\end{center}
\end{figure}

In practical calculations, we set an electron with a spin $\alpha$ initially in the ground state of the harmonic confinement $\omega_0=15$ [a.u.]. Then, we applied an external Rashba field perpendicular to the 2D plane with resonant frequency $w=\omega_0$, which is equal to the energy difference between the eigenstates.
Figure \ref{snap} shows snapshots of the $\alpha$ and $\beta$ spin densities with respect to time. 
The $\alpha$ spin decreases, whereas the $\beta$ spin increases and becomes recognizable when $t \sim 2$ [a.u.]. The generated $\beta$ spin distribution has a single node in the radial direction.
When time passes $t \sim 4$, the $\alpha$ spin disappears and the $\beta$ spin becomes dominant. Subsequently, the $\beta$ spin decreased, whereas the $\alpha$ spin increased.
Finally, the system completely restores the initial spin distribution at $t \sim 8$. These TD features were repeated periodically.

\subsection{Multiple and sequential transitions}
\label{statepro}

Because we neglect $\bm l \cdot \bm s$ SOI coupling in the unperturbed Hamiltonian of Eq. \eqref{H0}, the space-eigenstate $\ket{nl}$ and spin-eigenstate $\ket{\sigma}$ are independent. 
We express the unperturbed state $\ket{nl\sigma} \equiv \ket{nl}\cdot \ket{\sigma}$ by $(nl\sigma)$, and numerically project the calculated TD wavefunction $\ket{\psi(t)}$ onto these eigenstates in accordance with $\sigma$:
\begin{align}
\label{pj0}
\ket{\psi(t)}  = 
\sum_{nl} C_{nl\alpha}(t) \ket{nl\alpha} + \sum_{nl}C_{nl\beta}(t) \ket{nl\beta}. 
\end{align}
Here, $C_{nl\sigma}(t)=\braket{nl\sigma | \psi(t)}$ denotes the expansion coefficient projected onto the unperturbed state $(nl\sigma)$.
The probability of each state against time was then calculated as $P_{nl\sigma}(t)=|C_{nl\sigma}(t)|^2$.

\begin{figure}[hbtp]
\begin{center}
\includegraphics[scale=0.6]{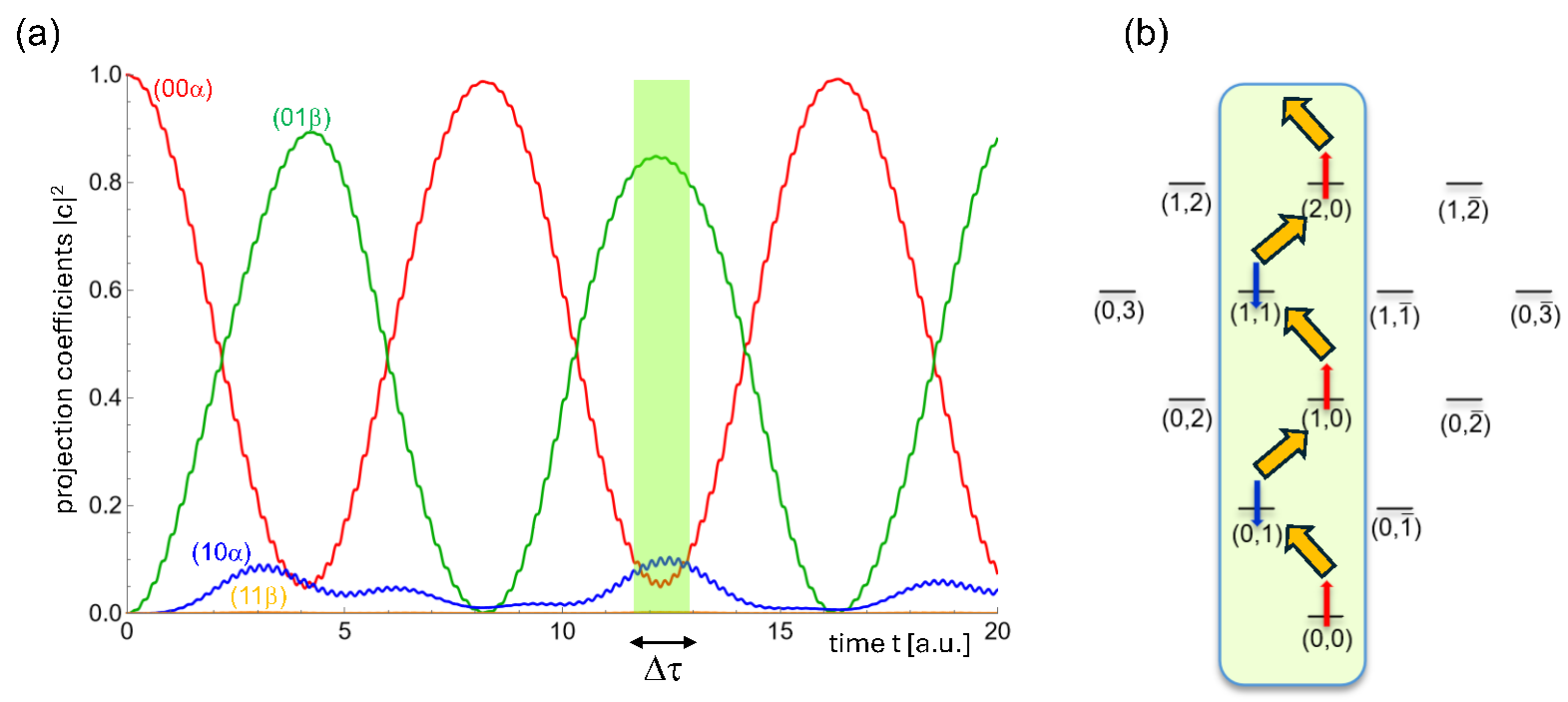}
\caption{State probability against time (a) and possible transitions driven by the alternating Rashba field in harmonic confinement. (b) Note that during $\Delta \tau$ as shown in figure (a), the component $\alpha$ is caused mainly by the state $(10\alpha)$ rather than by the state $(00\alpha)$.}
\label{fig3}
\end{center}
\end{figure}

Figure \ref{fig3}(a) presents the state probability with respect to time. The electron is initially in the ground state ($00\alpha$) with spin $\alpha$ and then transitions to the first excited state $(01\beta)$. 
However, even the maximum probability of state $(01\beta)$ is less than $90 \%$, and is not full. 
Projection analysis revealed that the remaining $10\%$ was shared by the higher excited states of $(10\alpha)$ and $(11\beta)$. 
Thus, the Rashba SOI in harmonic confinement causes multiple excitations and successive transitions to higher excited states. 
Figure \ref{fig3}(a) further demonstrates that the electrons that dissipated into the higher states returned to the original ground state after $t \sim 17$, repeating this periodically.

Because confinement is achieved by the 2D central force field, the interstate SOI transition occurs between states with different angular momenta ($l$ and $l^\prime$) owing to the in-plane momentum operators in Eq. \eqref{Fdef}. 
Furthermore, the conservation of the total angular momentum $j_z=l+s_z$ controls the present Rashba SOI selection rule, because the Rashba field is applied perpendicular to the 2D plane \cite{jjap1,jjap2,inui}. Eventually, the interstate Rashba SOI coupling is given by
\begin{equation}
\label{seni3}
\langle n^\prime l^\prime \sigma^\prime | \hat{\mathcal H}_R | n l \sigma \rangle 
\propto \delta_{j_z^\prime, j_z}\times   \delta_{n^\prime, n} \left \{
\begin{array}{l}
  {\rm or} \  \times \delta_{n^\prime, n-1} \ (\alpha), \\
  {\rm or} \ \times  \delta_{n^\prime, n+1} \ (\beta).
\end{array}
\right.
\end{equation}

Figure \ref{fig3}(b) illustrates the possible interstate transitions caused by Rashba SOI. Harmonic confinement causes multiple but inter-neighboring-state couplings, and the conservation of $j_z$ changes the spin state alternately.
Eventually, the ``multiple but successive" interstate couplings results with the spin flip-flopping, as illustrated in Fig. \ref{fig3}(b). The meaning of the ``sequential" couplings is naturally symbolic and has the actual time-delays.

\subsection{Rabi frequency in the harmonic confinement}
\label{rabibi}

An ideal Rabi oscillation occurs resonantly between the two states of the initial (ground $\ket{n}$) and the final (excited $\ket{k}$) states.
In this simple interstate transition, the rotating wave approximation determines the state probability via the Rabi frequency $\Omega_\mathrm R$ as follows:
\begin{eqnarray}
P_i(t) =|C_i(t)|^2 \propto \sin^2 \left(\frac{\Omega_\mathrm R}{2} t\right), \ \  (i = n \ {\rm or} \ k). 
\end{eqnarray}
Furthermore, by adding the rotating-wave approach to the two states consideration (TSRW), we obtain the Rabi frequency as: 
\begin{eqnarray}
\Omega_\mathrm R ^2= \frac{4}{\hbar^2}|F_{kn}|^2.
\end{eqnarray} 
The interstate transition matrix element is defined as $F_{kn}$: 
For the proposed Rashba SOI coupling (Eq. \eqref{Fdef}), this matrix element $F_{kn}$ is more explicitly given by,
\begin{eqnarray}
\label{Fkn}
F_{kn}=-\frac{1}{2i}\langle k | \hat{\mathcal H}_R | n \rangle =\frac{1}{2i} \Xi_0 \langle k | \hat{p}_y \sigma_x - \hat{p}_x \sigma_y | n \rangle.
\end{eqnarray}

Figure \ref{fig3}(a) shows that the primary transition occurred between the ground state $(00\alpha)$ and 1st excited state $(01\beta)$ and the obtained period was $T_{\mathrm R}=8.33$ [a.u.].
Therefore, a Rabi frequency of $\Omega_{\mathrm R}=2\pi/T_{\mathrm R}= 0.754$ was estimated.
By contrast, the transition matrix element is $|F_{ 01\beta,00\alpha}|=0.389$. 
Accordingly, the Rabi frequency was estimated ${\Omega_\mathrm R = \frac{2}{\hbar}|F_{ 01\beta,00\alpha}|= 0.778}$, which agrees with $0.754$.
Thus, the application of the TD Rashba field with a frequency of $w=\omega_0$ causes Rabi oscillations with a frequency of $\Omega_\mathrm R = \frac{2}{\hbar}|F_{ 01\beta,00\alpha}|$. 
The coupling of harmonic confinement with Rashba SOI results in a unique selection rule for the interstate transition.
Consequently, the Rabi oscillation results in multiple and successive ``stairway transitions."

\section{Bloch sphere representation}
\label{BSrep}

Projection analysis is useful to understand the physical phenomena in terms of the transition between unperturbed eigenstates. However, the calculation of the state probability removes the phase information accompanied by the wavefunction. Although the phases of individual wave functions are indefinable, the phase difference between two specific eigenstates is definable. Here, we extract the phase information found in the Rabi oscillation in harmonic confinement by rewriting the TD wavefunction as a BS.

\subsection{Zenith and azimuthal angles}

The BS consists of two mutually orthogonal and normalized bases of the north $\ket{N}$ and south $\ket{S}$ poles, respectively. 
Accordingly, the TD wavefunction $\ket{\varphi(t)}$ is expanded as
\begin{align}
  \ket{\varphi(t)}
   & = C_{N}{(t)}\ket{N} + C_{S}{(t)}\ket{S},
\label{eq:tdwave}
\end{align}
and complex expansion coefficients.

The TD wavefunction on BS is further determined by two BS parameters: the zenith angle $\theta_B$ and azimuthal angle $\phi_B$. 
The zenith angle $\theta_B$ indicates the mixing ratio between the two basis states and azimuthal angle $\phi_B$ indicates the phase difference between them.
By employing $\theta_B$ and $\phi_B$, the TD wavefunction $\ket{\varphi(t)}$ on BS is given by:
\begin{align}
  \ket{\varphi(t)} & =\mathrm{e}^{i\lambda_{\mathrm{B}}(t)}
  \left(
  \cos\frac{\theta_{\mathrm{B}}(t)}{2}
  \ket{N}
  +\mathrm{e}^{i\phi_{\mathrm{B}}(t)}
  \sin\frac{\theta_{\mathrm{B}}(t)}{2}
  \ket{S}
  \right),
  \label{blochsp}
\end{align}
defining the global phase as $\lambda_{\mathrm{B}}(t)$.

Consequently, when the TD wavefunction is obtained, the projection to the BS determines the two expansion coefficients $C_N(t)$ and $C_S(t)$. A comparison of Eq. \eqref{eq:tdwave} with Eq. \eqref{blochsp} (Appendix \ref{appBS}) gives the BS parameters, $\theta_B$ and $\phi_B$ by,
\begin{align}
\label{tbpb1}
\tB(t) &=2 \tan^{-1} \left( \frac{|C_{S}(t)|}{|C_N(t)|} \right), \nonumber \\
\phiB(t) & =\mathrm{Arg} \left( \frac{C_S(t)}{C_N(t)} \right).
\end{align}
Eventually, we can study the TD features of the hybridization and phase difference between the two bases using $\tB$ and $\phiB$, respectively.

\subsection{Primary transition}
\label{prim}

We now focus on the primary transition between $(00\alpha)$ and $(01\beta)$ by extracting the two states from the current Rabi oscillation (two-state extraction model).
Because we calculated the expansion coefficients $C_{00\alpha}(t)$ and $C_{01\beta}(t)$ numerically, we computationally determined the BS parameters $\tB(t)$ and $\phiB(t)$ using Eq. \eqref{tbpb1} as follows
\begin{align}
\label{tbpb0}
\tB(t) &= 2 \tan^{-1} \left(\frac{|C_{01\beta}(t)|}{|C_{00\alpha}(t)|}\right), \nonumber \\
\phiB(t)&=\mathrm{Arg}(C_{01\beta}(t))-\mathrm{Arg}(C_{00\alpha}(t)). 
\end{align}

\begin{figure}[htbp]
\begin{center}
\includegraphics[scale=0.6]{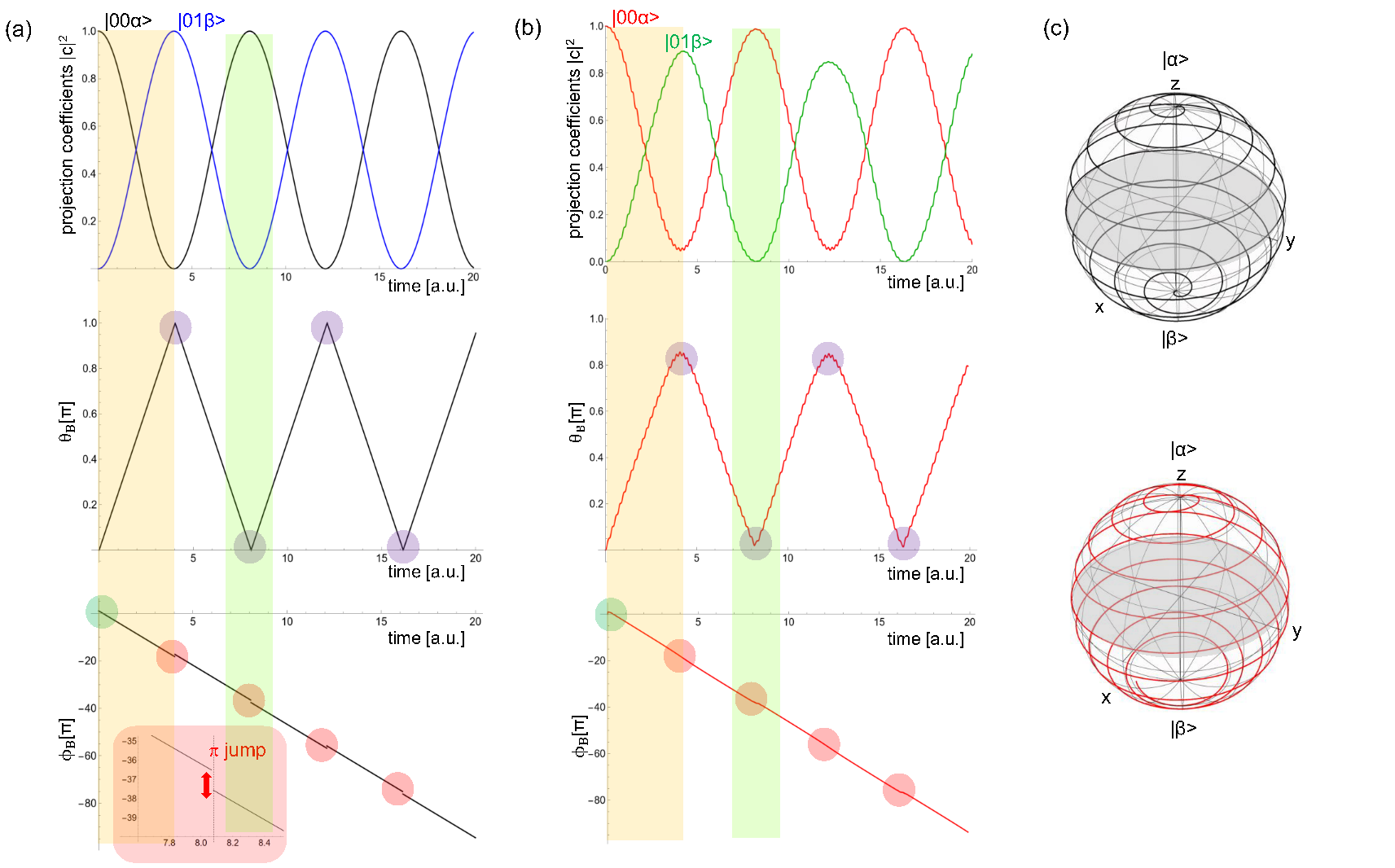}
\caption{Change in the state probabilities, BS parameters $\tB$ and $\phiB$ against time. Those for the TSRW approach are shown in figure (a) whereas those for the primary transition between $(00\alpha)$ and $(01\beta)$ are shown in figure (b). We further show each trajectory of the TD wavefunction during the single cyclic period (yellow region) in (c). Note that the trajectory for the TSRW approach passes exactly through the poles whereas that for the extract model only approaches the pole.
}
\label{grnd-1st}
\end{center}
\end{figure}

Figure \ref{grnd-1st}(a) shows that the zenith angle $\tB(t)$ forms a triangular wave over time. Moreover, it completely synchronizes with the projection coefficients, with the characteristic periodicity of the Rabi oscillation $\Omega_\mathrm R$. 
It should be noted that the extrema of the expansion coefficients occur when the TD wavefunction passes through each BS pole.
When the coefficient $C_{00\alpha}$ has a maximum value, the zenith angle $\tB$ has a minimum value of zero, causing an indifferentiable acute apex. 
In contrast, the zenith angle was maximized when the coefficient $C_{01\beta}$ was also maximized. However, the corresponding extremum does not amount to unity and the apex is obtuse and differentiable.
The eigenstates caused by harmonic confinement have regular energy intervals, $\hbar \omega_0$. Therefore, even the injection of a single resonant frequency $ w=\omega_0$ causes multiple successive transitions, dissipating the TD wavefunction into multiple eigenstates. Consequently, the state probability $|C_{01\beta}|^2$ is not unity and 
dissipation to other states causes an obtuse and differentiable apex formation.
Figure \ref{grnd-1st}(a) also demonstrates that $\phiB$ shows basically a linear dependence on time. However, the linearity is broken when the wavefunction passes around the BS pole, that is, near the apex of $\tB$. Furthermore, a finite intersection exists at $t=0$.

\subsubsection{TSRW approach}

To deepen our understanding of the TD behavior of BS parameters $\tB(t)$ and $\phiB(t)$, we assume that a Rabi oscillation occurs between the two specific states of the N and S poles.
Furthermore, the employment of the rotating wave (RW) approximation (Appendix \ref{appBS}) provides analytical solutions of the expansion coefficients $C_N$ and $C_S$ by 
\begin{align}
\label{CnCs0}
|C_N(t)|^2&=\cos ^2 (|F|t), \nonumber \\
|C_S(t)|^2&=\sin ^2 (|F|t).
\end{align}
Here, we define $F=F_{SN}/\hbar$. 
Accordingly, based on Eq. \eqref{tbpb1}, we have the BS parameters $\tB(t)$ and $\phiB(t)$ as, 
\begin{align}
\label{tbpb}
\tB(t)&=2\cos^{-1} \left(|\cos(|F|t)| \right), \nonumber \\ \phiB(t)&= -\omega_0 t - \frac{\pi}{2} \mathrm{Sign}[\tan (|F|t)]+\mathrm{Arg}(F) + 2n\pi. 
\end{align}

Equation \eqref{tbpb} demonstrates the characteristic features of the BS parameters $\tB(t)$ and $\phiB(t)$ in the full-TSRW system. 
The term $\cos^{-1} \left(|\cos(|F_{SN}|t)|\right)$ represents the $\tB(t)$ behavior, a triangular wave with a value of $0$ or $\pi$ at every apex. Moreover, at each apex, the wavefunction passes through the BS pole, and state is completely replaced by the opposite state. Accordingly, the triangular wave has a period of $T_\mathrm{R}={\pi}/{|F|}$, equal to the Rabi oscillation of $\Omega_\mathrm R=2|F|$.

Equation \eqref{tbpb} also reveals that the azimuthal angle $\phiB(t)$ exhibits the following three features in a full-TSRW system:
\begin{itemize}
\item $\phiB(t)$ has a linear dependence against time and its slope is caused by the difference between the dynamical terms; $\omega_{S}-\omega_{N}(=\omega_0$ in the harmonic confinement),
\item a discontinuity having a value of $\pi$ occurs when passing the BS pole; $\pi$ jump,
\item an interception at $t=0$ corresponds to an argument of the transition matrix element, causing a constant phase-shift in the azimuthal angle; $\mathrm{Arg}(F)$,
\end{itemize}

Figure \ref{grnd-1st}(b) shows the practical behavior of the BS parameters $\tB(t)$ and $\phiB(t)$ under the full-TSRW approximation, where the Rabi oscillation is assumed to occur fully between the two states $(00\alpha)$ and $(01\beta)$ by the alternating Rashba field with $w=\omega_0(=15)$.
As the zenith angle $\tB$ indicates the mixing ratio between these two states, its cycle synchronizes completely with that of the two-state coefficients, as shown in Fig. \ref{grnd-1st}(b). 
In the present harmonic confinement, the effective strength of the harmonic potential can be evaluated using the slope of the linear dependence. Figure \ref{grnd-1st}(b) shows that the slope of $15.0$ is equal to the confinement strength $E_{01}-E_{00}=\hbar \omega_0 =15$.

\subsubsection{Beyond TSRW}

\begin{figure}[htbp]
\begin{center}
\includegraphics[scale=0.59]{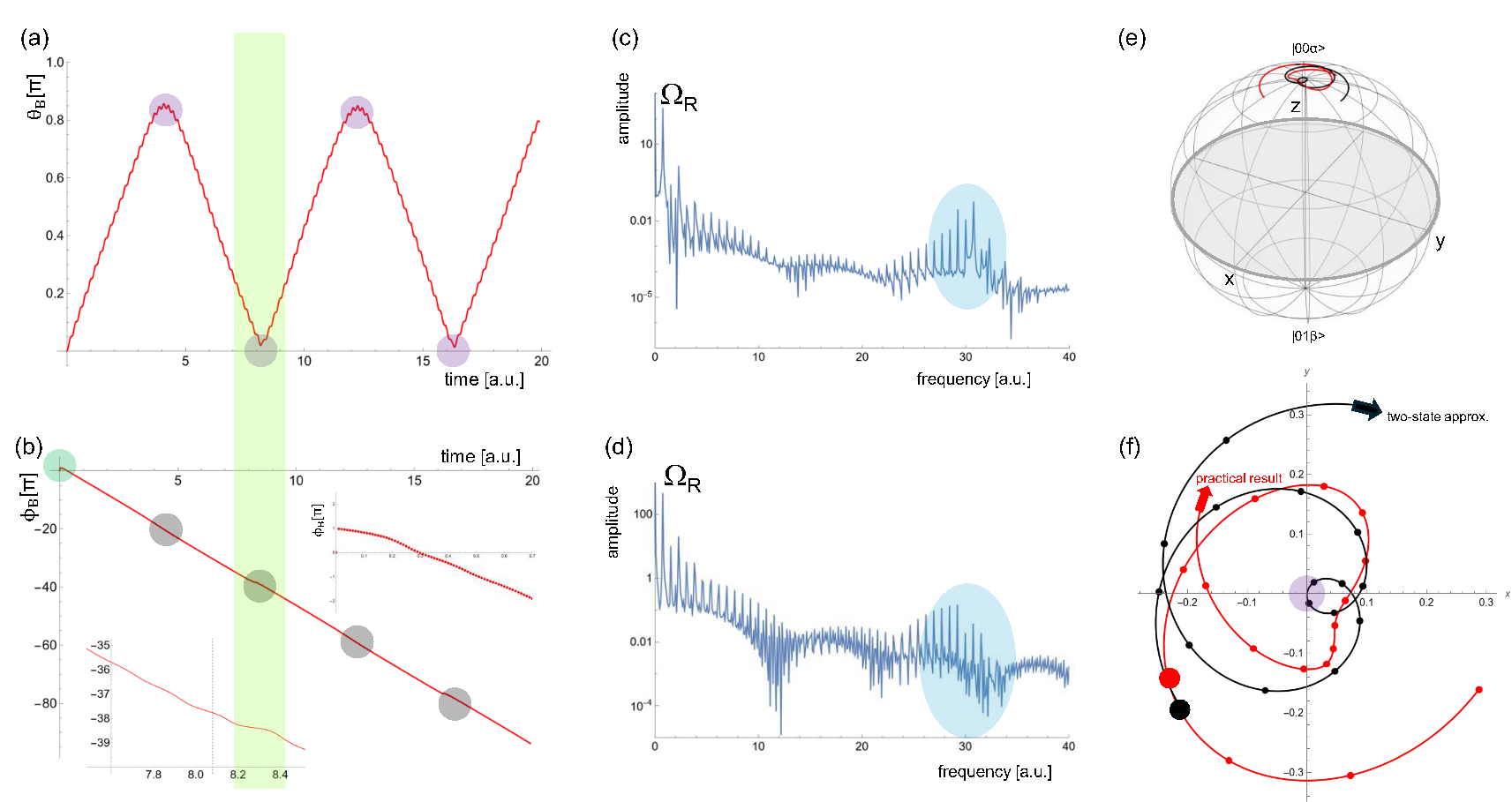}
\caption{Change in $\tB$ (a) and $\phiB$ (b) for the primary transition.
The intersection value $0.984\pi$ (right-inset of Figure (b)) is nearly equal to that by the full-TSRW approach ($\mathrm{Arg}(F)=\frac{3\pi}{2} - \frac{\pi}{2}=\pi$). The left-inset also shows the breaking of the linear dependence by the $\pi$ jump. In figures (c) and (d), we show the Fourier components of $\tB$ and $\phiB$, respectively. We further show the trajectory (e) and its projection onto the equatorial plane (f) during the specific time indicated by the green band. The trajectory by the TSRW (black line) passes definitely via the North pole whereas that for the practical primary transition (red line) only approaches the North pole.
}
\label{grnd-1stS}
\end{center}
\end{figure}

We return to the primary transition (Fig. \ref{grnd-1st}(a)) and study the details in $\tB(t)$ and $\phiB(t)$ by comparing with those with the full TSRW approach.
The calculated zenith angle exhibits a triangular wave against time, as predicted by the TSRW approach. 
According to Fig. \ref{grnd-1stS}, we have a cyclic period $T_{\mathrm R}=8.333$ of this triangle wave, resulting in a Rabi oscillation $\Omega_\mathrm R=2\pi/T_{\mathrm R}=0.754$.
However, the resulting apexes of $\tB(t)$ are obtuse and differentiable, particularly when passing through the BS south pole $\ket{01\beta}$. 
In contrast, the state at the N pole interchanges completely, leading to $\alpha$ spin. 
These complete and incomplete state interchanges deform the triangular wave into a sawlike one. 
The sawlike wave has $n$ times the harmonic components of the specific frequency $\omega_s$, where $n$ is an integer. 
The Fourier transform clearly revealed the existence of characteristic higher components (Fig. \ref{grnd-1stS}(b)). They are equal to the $n$-th harmonic Rabi components with integer number $n$, and the specific frequency is the Rabi oscillation $\Omega_\mathrm R$. 
The approach beyond the RW approximation predicts the existence of the twice-as-large (second harmonic) component $2\omega_0$ and further to be split into $2\omega_0 \pm \Omega_\mathrm R$ by the Rabi oscillation (Appendix \ref{bunretsu}). However, the numerous harmonic Rabi components obscure the separation by the Rabi oscillation $\pm \Omega_\mathrm R$ at the second-harmonic frequency $2\omega_0$ (Fig. \ref{grnd-1stS}).

Figure \ref{grnd-1stS} further demonstrates that the calculated $\phiB$ exhibits a linear dependence on time, as predicted by the TSRW approach. 
The obtained slope of $15.0$ coincides with the difference in the dynamical terms between the ground and 1st excited states, which is $\omega_0=15$ in the present harmonic confinement. 
The linear dependence was also broken at the passing pole, and the initial intersection of $\pi$ was observed (inset of Fig. \ref{grnd-1st}).
Thus, this two-state extracted model has three characteristic features as found in the TSRW approach.
However, the Fourier analysis elucidates the existence of higher oscillating components. The typical $\pi$ jumps found in the TSRW approach were also modified in the extracted model.

\subsubsection{$\pi$ jump}

\begin{figure}[htbp]
\begin{center}
\includegraphics[scale=0.5]{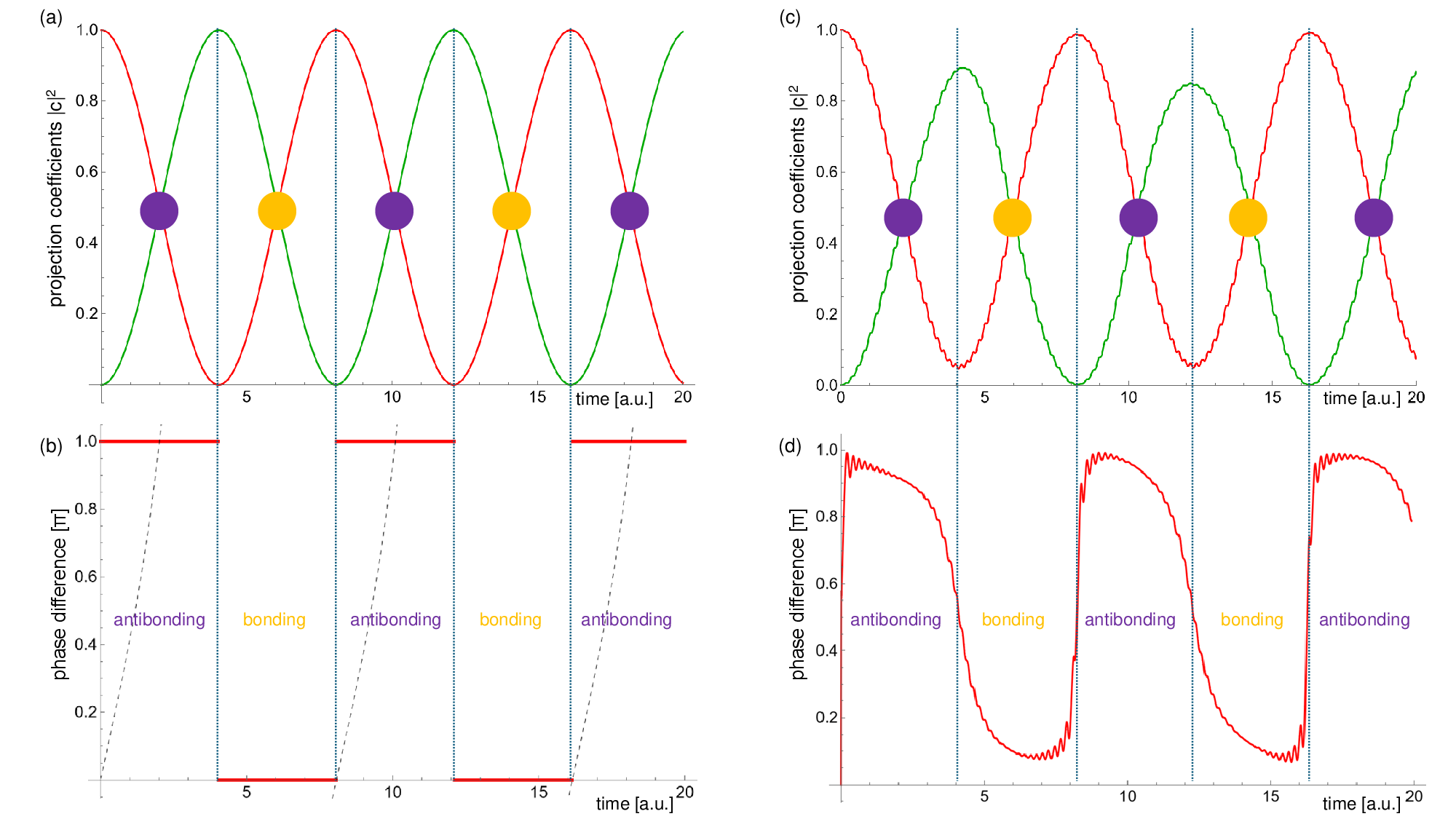}
\caption{Comparison between the expanding coefficients and the subtracted azimuthal angle for the TSRW approach (a, b) and practical primary transition (c, d).}
\label{step}
\end{center}
\end{figure}

The removal of the linear time development ($-\omega_0 t$) from $\phiB(t)$ implies the elimination of the dynamical terms from the system, leading to the extraction of a phase excluding the dynamical terms. In the full TSRW approach, this removal changes $\phiB$ into a step-like behavior between $0$ and $\pi$, leading to a square wave, as shown in Fig. \ref{step}(a). This characteristic feature originates from the ``$\pi$ jump" at the BS-pole passing.
The TSRW wavefunction propagated on the BS surface and the BS parameters varied smoothly until the wavefunction reached the BS pole. 
However, the BS pole has a {\it singularity}. 
Therefore, only when the TSRW wavefunction passes through the BS pole, the wavefunction suddenly but inevitably moves out to the other ``meridian section" of the BS, that is, a phase difference of $\pi$ is generated at the azimuthal angle. 
This feature is similar to Wayfarer's sudden encounter in his crossing over a pole, walking from the Western Hemisphere to the Eastern Hemisphere, and vice versa, except when he walks along the Greenwich meridian line. 
Accordingly, in the TSRW Rabi oscillation driven by the alternating Rashba field, we can find the phase differences $\pi$ and $2n\pi(=0)$, which emerge alternately when the wavefunction passes the BS pole.

During the first period, until the full-TSRW wavefunction is fully replaced by the base of the S-pole (Fig. \ref{step}(a) and (b)), the S-pole component ($e^{i\phi_S(t)}$) has a phase different than that of the N-pole component ($e^{i\phi_N(t)}$) by $\pi$, as described in Appendix \ref{appBS}. 
Subsequently, these two bases have the same phase $0(+2n\pi)$ during the second period until the TD wavefunction is fully restored by the N-pole base.
It should be noted that the wavefunction is at the crossing point of the projection coefficients, where the two pole states are mixed equally. In the first period, the phase difference of $\pi$ causes an ``antibonding" character between the two pole-states whereas the zero phase difference causes a ``bonding" one in the second period. 
Thus, the BS representation illustrates the periodic appearance of the bonding and antibonding states between the two poles, resembling an emerging ``Floquet" state.
The $\pi$ jump might be a difference between ``topological" phases in the BS space. 
However, the net phase difference disappears when $\pi+\pi$, when the cycle period returns fully.
This is because the present application of the alternating Rashba field varies only the value $\Xi_0$ within $\pm|\Xi_0|$, that is, the one-dimensional change in the system parameter. 
Consequently, a single cycle of a $\Xi_0$ change only results in a value of $2\pi$.

Figures \ref{step}(c) and (d) demonstrate that $\phiB$ in the primary transition preserves the step-like feature between $0$ and $\pi$, synchronizing with the change in the projection coefficients.
The multiple and successive stairway transitions caused by the Rabi oscillation result in an incomplete transition to the S-pole state $(01\beta)$. Consequently, the step edge of $\phiB$ is chipped off, and a deformed square wave results. 
Eventually, the $\phiB$ discontinuity caused by the $\pi$ jump disappears, and the continuity is restored by a beat wave with higher-frequency components.
The Fourier investigation shows that the modified square wave comprises three types of oscillation components: the second harmonic of the resonant frequency $2\omega_0$ and the $n$-th ($n$: integer) harmonic components of the Rabi oscillation $\Omega_\mathrm R$. 
The even-numbered harmonic components were caused by sawlike waves, whereas the odd-numbered components were caused by the superposition of the square and sawlike waves.

\subsection{Individual transition processes}

As predicted by the selection rules in Eq. \eqref{seni3}, Figure \ref{fig3} indicates the existence of the other transitions in addition to the primary transition between $(00\alpha)$ and $(01\beta)$. These are the secondary transitions between $(01\beta)$ and $(10\alpha)$, the third between $(10\alpha)$ and $(11\beta)$, etc.
Similar to the primary transition, we extract each transition and represent it individually using the corresponding BS.

\begin{figure}[htbp]
\begin{center}
\includegraphics[scale=0.25]{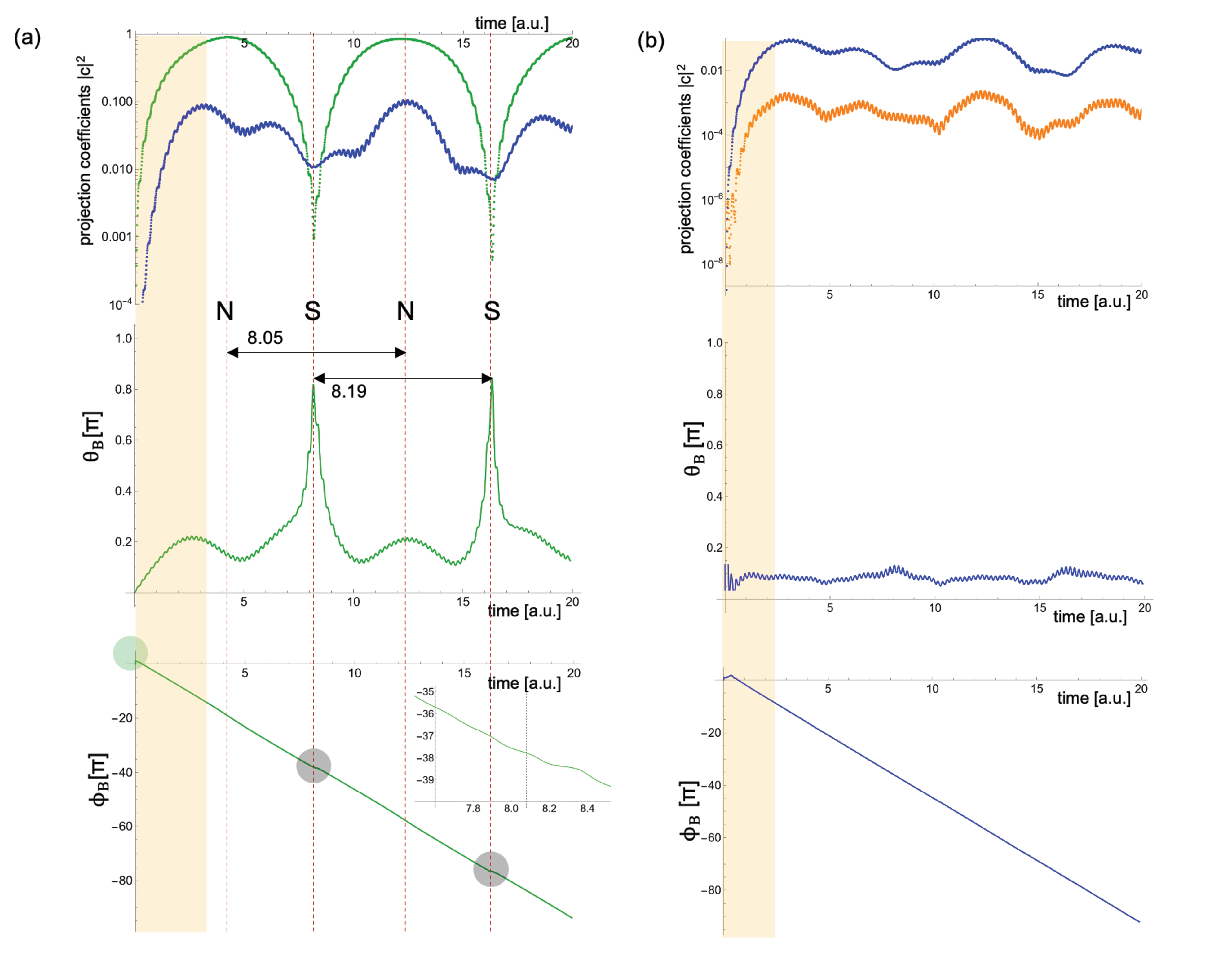}
\caption{Change in the state probabilities, BS parameters $\tB$, and $\phiB$ against time for the secondary (a) and tertiary transition (b).}
\label{multi}
\end{center}
\end{figure}

Fourier investigation of the secondary transition revealed two characteristic frequencies: Rabi ($\Omega_\mathrm R \sim 0.754$) and the second harmonic ($\omega_0 + w \sim 29.22$). 
The zenith angle $\tB$ no more, however, exhibits a well-defined triangular wave with respect to the time, 
because a small expansion coefficient $C_{10\alpha}$ breaks the validity of the two-state approach.
It rather has a spike peak around $8$ and $16$ [a.u.], which obscures the periodicity of $\tB$, except for the Rabi period between ``spikes."
Considering Eq. \eqref{tbpb1}, these spikes are caused by the extreme reduction in the denominator $C_{01\beta}$ when the TD wavefunction passes/approaches the N-pole of $(01\beta)$ (Fig. \ref{grnd-1stS}(e) and (f)). 
Because the N-pole state of $(01\beta)$ is initially empty, the time period (8.05) until the next N-pole is slightly different from that between the inter-S-poles (8.19). 
By contrast, the azimuthal angle $\phiB(t)$ exhibits a linear dependence on time (Figure \ref{multi}(a)).
However, the TD wavefunction in this simple two-state extraction model does not pass the pole states and produces indistinct $\pi$ jump results, as shown in the inset of Fig. \ref{multi}(a).

Figure \ref{multi}(b) shows the TD features of BS parameters $\tB(t)$ and $\phiB(t)$ for the third transition.
The extremely small transition obscures the characteristic triangular wave of $\tB(t)$ whereas the linear dependence of $\phiB(t)$ remains. However, a distinct $\pi$ jump is no longer recognizable.

\section{Effective Bloch Sphere}
\label{seceffBS}

The Rabi oscillation in the harmonic confinement causes multiple and successive ``stairway transitions." Therefore, each transition requires a corresponding BS representation as described previously.
However, the snapshots in Fig. \ref{snap} demonstrate that the alternating Rashba field having a resonant frequency $w=\omega_0$, which changes the TD spin state alternately and periodically between $\alpha$ and $\beta$.
By focusing on the selection rules in Eq. \eqref{seni3}, all transitions can be contracted into the change in the spin state between $\alpha$ and $\beta$.
Thus, by setting the two BS poles based on $\ket{\alpha}$ and $\ket{\beta}$, we can effectively rewrite the individual transitions as a two-state transition between $\ket{\alpha}$ and $\ket{\beta}$. 
Eventually, we transform the multiple and successive ``stairway transitions" into a single BS, and discuss practically the time development of both the mixing ratio and phase difference between these two spin states.

Now, we introduce the contracted expansion coefficients $C^\alpha (t)$ and $C^\beta (t)$, leading to the ``spinor" description of the system.
Equation \eqref{pj0} defines the coefficients $C^\alpha$ and $C^\beta$ as follows:
\begin{align}
\ket{\psi(t)}
&=\sum_{nl} C_{nl\alpha}(t) \ket{nl}\cdot\ket{\alpha} + \sum_{nl}C_{nl\beta}(t) \ket{nl}|\cdot\ket{\beta} \nonumber \\
&\equiv C^\alpha(t) \ket{\varphi_\alpha(\bm r)}\cdot \ket{\alpha}+C^\beta(t) \ket{\varphi_\beta(\bm r)}\cdot\ket{\beta}.
\end{align}
Here, $C_{nl\sigma}(t)$ is the TD expansion coefficient projected onto the unperturbed state ($nl$) with spin $\sigma$.
Accordingly, the space-dependent terms $\ket{\varphi_\alpha(\bm r)}$ and $\ket{\varphi_\beta(\bm r)}$ are defined as
\begin{align}
\ket{\varphi_\alpha(\bm r)}&=\frac{1}{C^\alpha(t)}\sum_{nl} C_{nl\alpha}(t) \ket{nl}, \nonumber \\
\ket{\varphi_\beta(\bm r)}&=\frac{1}{C^\beta(t)}\sum_{nl} C_{nl\beta}(t) \ket{nl}.
\end{align}
Normalization $\braket{\psi(t) | \psi(t)}=1$ yields the following:
\begin{align}
\label{C2}
|C^\alpha(t)|^2=\sum_{nl}|C_{nl\alpha}(t)|^2, \nonumber \\
|C^\beta(t)|^2=\sum_{nl}|C_{nl\beta}(t)|^2.
\end{align}
Note that each contracted coefficient, $C^\alpha(t)$ and $C^\beta(t)$, inevitably has phase indefiniteness $e^{i \zeta_\alpha}$ and $e^{i \zeta_\beta}$, respectively. Accordingly, we obtain the contracted coefficients as
\begin{align}
\label{Cabs}
C^\alpha(t)=e^{i \zeta_\alpha}\sqrt{\sum_{nl}|C_{nl\alpha}(t)|^2}, \nonumber \\
C^\beta(t)=e^{i \zeta_\beta}\sqrt{\sum_{nl}|C_{nl\beta}(t)|^2}.
\end{align}
Here, we approximate the effective phase $\zeta_\sigma$ by weight averaging the phase $\phi_{nl\sigma}$ in the individual process by
\begin{align}
\zeta_\sigma &= \mathrm{Arg} \displaystyle{\left( \sum_{nl} e^{i \phi_{nl\sigma}} |C_{nl\sigma}|^2 \right) }.
\end{align}

\begin{figure}[htbp]
\begin{center}
\includegraphics[scale=0.5]{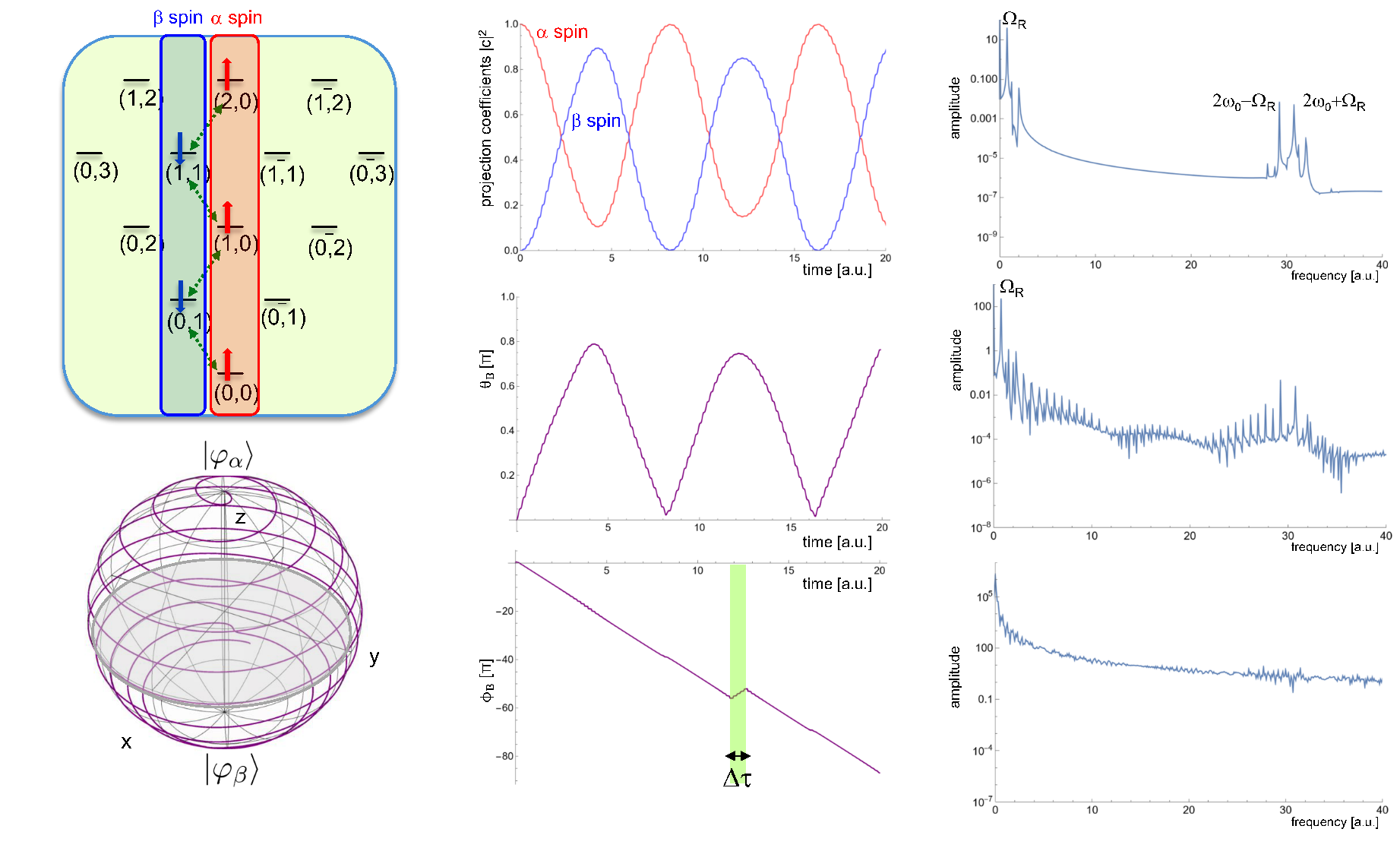}
\caption{Effective BS representation approach; possible transition driven by the alternating Rashba field and resulting trajectory on the BS. We also show the change in $\tB^{\mathrm{eff}}$ and $\phiB^{\mathrm{eff}}$ for the effective BS presentation with the effective coefficients $C^\alpha(t)$ and $C^\beta(t)$. 
Fourier analysis elucidates the characteristic components for $C^\alpha(t)$, $\tB^{\mathrm{eff}}$, and $\phiB^{\mathrm{eff}}$. Breaking of the linear dependence in $\phiB^{\mathrm{eff}}$ is highlighted by green circles.}
\label{effBS}
\end{center}
\end{figure}

Similar to the two-state extracted model above, the effective zenith angle 
$\tB^\mathrm{eff}= 2 \tan^{-1} {\left({|C^\beta|}/{|C^\alpha|}\right)}$ 
also exhibited a triangular wave with respect to time. 
The spin interchange at the S-pole is incomplete, and the state does not change fully into $\beta$ whereas the state at the N-pole completely interchanges, leading to $\alpha$ spin (Fig. \ref{effBS}).
Thus, this incomplete spin interchange causes the non-acute apexes of the ``triangle" wave every when the TD wavefunction passes the S-pole, and the resulting $\tB^\mathrm{eff}$ does not attain $\pi$.
Nevertheless, the deformed triangular wave had a primary period approximately equal to the Rabi period of $\Omega_R=2|F|=0.754$. 
Fourier analysis further reveals the existence of other higher components equal to the $n$-th harmonic Rabi components ($n \times \Omega_\mathrm R$) with an integer number $n$, as discussed in the two-state extracted model.
Furthermore, these numerous components similarly obscure the separation by the Rabi oscillation $\pm \Omega_\mathrm R$ at the second-harmonic frequency $2\omega_0$.

An approximately linear dependence was also observed for the effective azimuthal angle $\phiB^{\mathrm{eff}}$ with respect to time.
As discussed in Subsec. \ref{prim}, the slope against time is equal to the difference of the dynamical terms of the two states.
The resulting slope of $\phiB^{\mathrm{eff}}$ is $-15.04 \pm 0.01$ near the harmonic confinement $\omega_0=15.0$.
Therefore, the harmonic potential ``equally" confines the two spin states.

The reversal of the slope around $t\sim 12$ is of interest, which leads to a positive slope of $12.27 \pm 0.47$. This feature originates from multiple successive Rabi transitions under harmonic confinement.
Figure \ref{fig3} demonstrates that the $\alpha$ spin around $t\sim 12$ is caused by the $(10\alpha)$ state rather than the $(00\alpha)$ state, whereas the $(00\alpha)$ state is dominant in other periods. This also reveals that the $\beta$ spin is always caused by the $(01\beta)$ state.
It should be noted that the slope of the azimuthal angle $\phiB$ against time is proportional to the difference $\Delta E=-(E_S-E_N)$ between eigenvalues $E_N$ and $E_S$. 
During this period, with the exception of $\Delta \tau$ in Fig. \ref{fig3}, $\Delta E=-(E_{01\beta} -E_{00\alpha})= -(2\hbar \omega_0 - \hbar \omega_0)= -\hbar \omega_0$.
By contrast, around $t\sim 12$,  $\Delta E=-(E_{01\beta} -E_{10\alpha})= -(2\hbar \omega_0 - 3\hbar \omega_0 )= \hbar \omega_0$, leading to the elucidation of the slope-sign reversion around $t\sim 12$.
Thus, slope analysis of the effective azimuthal angle $\phi_B^{\mathrm {eff}}$ elucidates the dominant transition.

\begin{figure}[htbp]
\begin{center}
\includegraphics[scale=0.7]{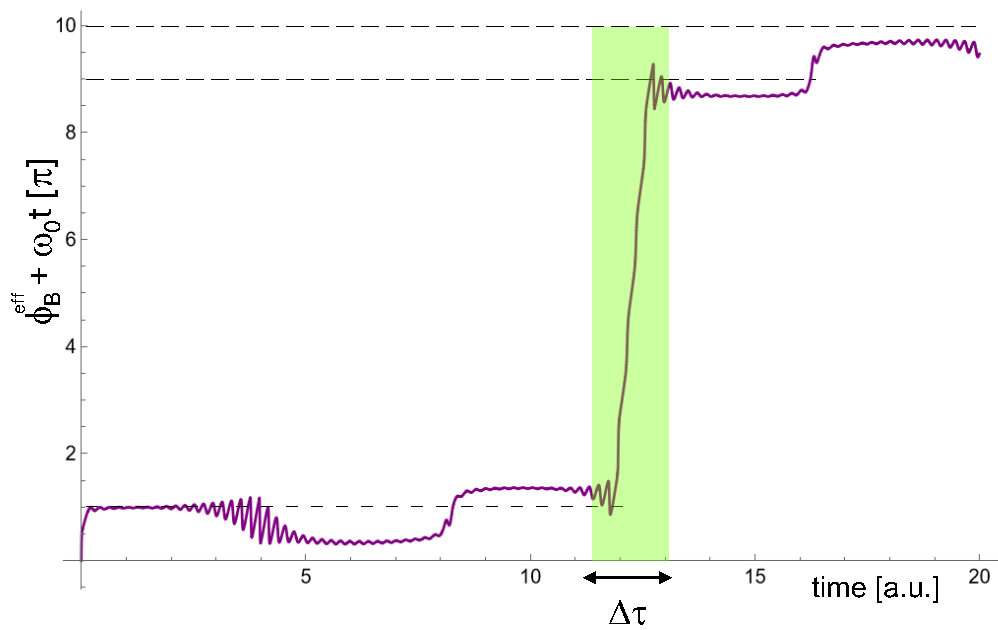}
\caption{Change in the effective azimuthal angle $\phiB^{\mathrm{eff}}$ against time by subtracting the linear dependence. Note that the jump during $\Delta \tau$ is caused by the change in the dominant $\alpha$ state (see Fig. \ref{fig3}), as described in text.}
\label{effPi}
\end{center}
\end{figure}

Furthermore, we subtract the linear TD term $(-\omega_0 t)$ from the effective azimuthal angle $\phiB^{\mathrm{eff}}(t)$ and deepen our understanding of the $\pi$ jump in effective BS representation (Fig. \ref{effPi}).
A comparison of the results shown in Fig. \ref{step} reveals that the effective BS representation has a more deformed square wave.
The step-edge deformation was enhanced by the chipping caused by the superposition of the high-frequency components, particularly when passing the S-pole. 
The Rabi oscillation in harmonic confinement disperses electrons into numerous states according to the selection rule Eq. \eqref{seni3}. 
Consequently, a state with complete spin $\beta$ does not appear if the spin $\alpha$ is initially set (Fig. \ref{fig3}).
The two-state extraction model partially illustrates this incompleteness, whereas the effective BS representation fully describes it.

\section{Summary}

We study the dynamical properties of ``multiple but successive" Rabi oscillations driven by the alternating Rashba field applied to the 2D harmonic confinement system. 
We solved the TD Schr\"{o}dinger equation and rewrote the TD wavefunction onto the BS by employing the two parameters of the zenith angle $\theta_B$ and azimuthal angle $\phi_B$.
The zenith angle $\theta_B$ indicates the mixing ratio between the two BS-pole states whereas the azimuthal angle $\phi_B$ indicates the phase difference between them.
Thus, the BS representation is particularly useful for the extraction of the phase information of the system.

The TSRW approach reveals the fundamental TD features of $\theta_B$ and $\phi_B$; the former causes a triangular wave, whereas the latter changes linearly with time. 
At each apex of the triangular wave, the wavefunction passes through the BS pole and the state is completely replaced by an opposite spin state. 
The slope of the linear change in $\phi_B$ is equal to the difference between the dynamical terms, leading to a confinement potential in harmonic confinement. Furthermore, the TSRW approach elucidates the existence of phase jumping by $\pi$, when the wavefunction passes through the BS pole.
The TSRW approach is not strictly applicable to multiple successive Rabi oscillations in harmonic-confinement systems. 
Nevertheless, we can find the triangular waveform of $\tB$, linear dependence, and $\pi$ jump in $\phiB$ even in complicated Rabi oscillations beyond the TSRW approach.

In this paper, we propose an EBS representation by transforming multiple successive transitions onto a BS sphere with poles $\alpha$ and $\beta$.
The multiple Rabi transition disperses the electron into numerous states; therefore, a state with complete spin $\beta$/$\alpha$ does not appear if the spin $\alpha$/$\beta$ is initially set. 
Therefore, the individually extracted two-state model did not illustrate this incompleteness.
However, this EBS representation effectively demonstrates the practical TD features of the mixing and phase difference between the two spin states, irrespective of multiple and successive transitions.   
The combination of the mixing ratio $\tB^{\mathrm {eff}}$ and slope analysis of $\phiB^{\mathrm {eff}}$ elucidates the dominant transition in multiple Rabi oscillations. Furthermore, the subtraction of the linear dependence from $\phiB^{\mathrm {eff}}$ reveals a phase difference of $\pi$ when the TD wavefunction passes around the poles.

\section*{Acknowledgements}
This is a part of the outcome of research performed under a Waseda University Grant
for Special Research Projects (Project numbers: 2023C-162 and 2024C-486).

\appendix

\section{Calculational details}
\label{cal}

\subsection{Real-space and -time difference method}
\label{RST}

The wavefunction is discretized into square grid points in real space (that is, a numerically exact diagonalization). The second(2nd)-order differential in the kinetic energy was determined using the central difference method. We employed Cartesian $x$ and $y$ grids by dividing the real space into a mesh of $64\times 64$ (1.538 nm squares). We then perform numerical diagonalization in a two-fold larger space owing to the spin hybridization between the spin-up ($\alpha$) and spin-down ($\beta$) states. Consequently, the eigenstates are doubly degenerated by $\alpha$ and $\beta$. Because the present 2D QD is supposed to be fabricated using InSb \cite{Winkler2000}, an effective mass (ratio) $m^\ast/m_0$ = 0.0138 \cite{koteles} and a dielectric constant (ratio) $\epsilon^\ast/\epsilon_0$ =17.88 \cite{young, adachi} of InSb are employed in the numerical calculation.

\subsection{Fourth-order Runge-Kutta time-developing method}
\label{4RK}

The TD Schr\"{o}dinger equation \eqref{td1} is given by
\begin{align}
\label{gene}
\frac{d}{d t}\ket{\psi(t)}
&=\frac{1}{i\hbar} \left(\hat{\mathcal H}_0(\bm r) + \hat{\mathcal H}_R(t) \right) \ket{\psi (t)} \nonumber \\
&\equiv {\bf g} (t, \ket{\psi(t)}).
\end{align}
The 4th-order Runge-Kutta method is defined as
\begin{align}
\ket{s_1}&={\bf g} (t, \ket{\psi(t)}), \nonumber \\
\ket{s_2}&={\bf g} (t+\frac{\Delta t}{2}, \ket{\psi(t)}+\frac{\Delta t}{2}\ket{s_1}), \nonumber \\
\ket{s_3}&={\bf g} (t+\frac{\Delta t}{2}, \ket{\psi(t)}+\frac{\Delta t}{2}\ket{s_2}), \nonumber \\
\ket{s_4}&={\bf g} (t+\Delta t, \ket{\psi(t)}+\Delta t\ket{s_3}).
\end{align}
Therefore, we can determine the TD wavefunction after the time evolution $\Delta t$ by
\begin{align}
\label{gene2}
\ket{\psi(t+ \Delta t)} \sim
\ket{\psi(t)} + \frac{\Delta t}{6} (\ket{s_1}+\ket{s_2}+\ket{s_3}+\ket{s_4}).
\end{align}

\section{Rate equation and TSRW approximation}
\label{fTSRW}

\subsection{Rate equation for two states}

To deepen our understanding of the TD dependence of the BS parameters $\tB(t)$ and $\phiB(t)$, we assume that the Rabi oscillation occurs between the two specific states because the BS representation is most appropriate when the TD wavefunction is fully expressed by the two states of the BS poles. For the primary transition, the two states are $\ket{N}=\ket{00\alpha}$ and $\ket{S}=\ket{01\beta}$.
Using the expansion coefficients $C_N$ and $C_S$, we first formulate $\tB$ and $\phiB$ under the assumption of two-state oscillation.
The corresponding coefficients are determined using the following coupled rate equation:
\begin{align}
\label{TSRW1}
i\hbar \frac{d}{dt} C_N &= \left[F_{NN} \mathrm e^{-iwt} + (F_{NN})^\ast \mathrm e^{+iwt} \right] C_N \nonumber \\
&+\left[ \mathrm e^{-i(\omega_S -\omega_N +w)t}F_{NS} 
+\underline{\mathrm e^{-i(\omega_S -\omega_N -w)t}(F_{SN})^\ast} \right] C_S, \nonumber \\
i\hbar \frac{d}{dt} C_S &= \left[ F_{SS} \mathrm e^{-iwt} + (F_{SS})^\ast \mathrm e^{+iwt}\right] C_S \nonumber \\
&+\left[ \underline{\mathrm e^{-i(\omega_N -\omega_S +w)t}F_{SN}} 
+\mathrm e^{-i(\omega_N -\omega_S -w)t}(F_{NS})^\ast \right] C_N.
\end{align}
Here, we remove each dynamical term from each expansion coefficient and define the eigen frequencies of eigenstates $\ket{N}$ and $\ket{S}$ as $\omega_N$ and $\omega_S$. The perturbed Rashba operator $\hat{F}$ in Eq. \eqref{Fdef} results in the transition matrix element $F_{SN}$ between those states.

\subsection{Rotating wave approximation in the harmonic confinement}

Harmonic confinement results in a distinct angular number $l$ in the nonperturbed state $(nl\sigma) (=\ket{nl}\cdot \ket{\sigma})$.
Because the Rashba perturbation has momentum and spin operators, the self-transition matrix elements $F_{NN}$ and $F_{SS}$ are removed and the first term on the right-hand side of Eq. \eqref{TSRW1} disappears.
We further employ the rotating wave (RW) approximation by which we drop the higher oscillating term ($\omega_S -\omega_N +w$) in Eq. \eqref{TSRW1}.
Accordingly, the TSRW approach for the 2D harmonic confinement simplifies the coupled rate equation \eqref{TSRW1} to
\begin{align}
\label{TSRW2}
i\hbar \frac{d}{dt} C_N &=  
\mathrm e^{-i(\omega_S -\omega_N -w)t}(F_{SN})^\ast C_S, \nonumber \\
i\hbar \frac{d}{dt} C_S &= 
\mathrm e^{-i(\omega_N -\omega_S +w)t}F_{SN} C_N.
\end{align}
As harmonic confinement results in $\omega_S -\omega_N=\omega_0$, Equation \eqref{TSRW2} is further simplified as
\begin{align}
\label{TSRW3}
i\hbar \frac{d}{dt} C_N &=  
\mathrm e^{-i(\omega_0 -w)t}(F_{SN})^\ast C_S, \nonumber \\
i\hbar \frac{d}{dt} C_S &= 
\mathrm e^{i(\omega_0 -w)t}F_{SN} C_N.
\end{align}
When the applied field has a resonant frequency of $\omega_0$, the coupled equations provide the analytical solutions for 
\begin{align}
\label{CnCs}
|C_N(t)|^2&=\cos ^2 (|F|t), \nonumber \\
|C_S(t)|^2&=\sin ^2 (|F|t).
\end{align}

\subsection{Splitting of the 2nd harmonic frequency}
\label{bunretsu}

Fourier analysis demonstrated splitting in the second-harmonic frequency, as shown in Fig. \ref{grnd-1stS}.
An approach beyond the rotating-wave approximation can explain this splitting. We return to Eq. \eqref{TSRW1}, where the transition is assumed only between the two states of N and S.
We solved this coupled equation semi-analytically using TSRW solutions.
According to Eq. \eqref{CnCs}, the two expansion coefficients $C_N(t)$ and $C_S(t)$ are given by,
\begin{align}
\label{CnCs2}
C_N(t)& \propto \cos \left(\frac{\Omega_\mathrm R}{2} t\right)
= \frac{1}{2}\left(\mathrm e^{i\frac{\Omega_\mathrm R}{2} t} +\mathrm e^{-i\frac{\Omega_\mathrm R}{2} t}\right), \nonumber \\
C_S(t)& \propto \sin \left(\frac{\Omega_\mathrm R}{2}t\right)
= \frac{1}{2i}\left(\mathrm e^{i\frac{\Omega_\mathrm R}{2} t} -\mathrm 
e^{-i\frac{\Omega_\mathrm R}{2} t}\right).
\end{align}
Consequently, the coupled equation \eqref{TSRW1} can be rewritten as follows:
\begin{align}
\label{TSRWd}
i\hbar \frac{d}{dt} C_N &\Rightarrow
\left[ \mathrm e^{-i(\omega_S -\omega_N +w)t}F_{NS} 
+{\mathrm e^{-i(\omega_S -\omega_N -w)t}(F_{SN})^\ast} \right] \frac{1}{2i}\left(\mathrm e^{i\frac{\Omega_\mathrm R}{2} t} -\mathrm e^{-i\frac{\Omega_\mathrm R}{2} t}\right), \nonumber \\
i\hbar \frac{d}{dt} C_S &\Rightarrow
\left[{\mathrm e^{-i(\omega_N -\omega_S +w)t}F_{SN}} 
+\mathrm e^{-i(\omega_N -\omega_S -w)t}(F_{NS})^\ast \right] \frac{1}{2}\left(\mathrm e^{i\frac{\Omega_\mathrm R}{2} t} +\mathrm e^{-i\frac{\Omega_\mathrm R}{2} t}\right).
\end{align}
Note that the relations between $\omega_S -\omega_N=\omega_0$, $(F_{SN})^\ast=-F_{NS}$, and the coupled equation \eqref{TSRWd} are simplified as
\begin{align}
\label{TSRWdd}
i\hbar \frac{d}{dt} C_N &\Rightarrow\frac{F_{NS}}{2i}\left(
\left[ \mathrm e^{-i(\omega_0 +w- \frac{\Omega_\mathrm R}{2})t}
-{\mathrm e^{-i(\omega_0 +w +\frac{\Omega_\mathrm R}{2})t}} \right]
-
\left[ \mathrm e^{-i(\omega_0 -w- \frac{\Omega_\mathrm R}{2})t}
-{\mathrm e^{-i(\omega_0 -w +\frac{\Omega_\mathrm R}{2})t}} \right]\right), \nonumber \\
i\hbar \frac{d}{dt}
C_S &\Rightarrow
\frac{(F_{NS})^*}{2}\left(
\left[ \mathrm e^{i(\omega_0 +w + \frac{\Omega_\mathrm R}{2})t}
+{\mathrm e^{i(\omega_0 +w -\frac{\Omega_\mathrm R}{2})t}} \right]
-
\left[ \mathrm e^{i(\omega_0 - w+ \frac{\Omega_\mathrm R}{2})t}
+{\mathrm e^{i(\omega_0 -w -\frac{\Omega_\mathrm R}{2})t}} \right]\right).
\end{align}
The expansion coefficients $C_N$ and $C_S$ have characteristic frequencies of $2\omega_0 \pm \frac{\Omega_\mathrm R}{2}$ when the Rashba field has a frequency of $w=\omega_0$.
As such, an approach beyond the RW approximation demonstrates that the 2nd harmonic frequency of $2\omega_0$ is split into $2\omega_0 \pm \Omega_\mathrm R$.

\section{BS parameters}
\label{appBS}

Here, we discuss the relationship between the expansion coefficients and the BS parameters.
The TD wavefunction is expressed by the following two methods employing the expansion coefficients and the BS representation:
\begin{align}
\ket{\psi(t)}&= C_N (t) \ket{N} + C_S (t) \ket{S} \nonumber \\
&= e^{i\phi_N(t)} \cos \frac{\tB(t)}{2} \ket{N} +e^{i\phi_S(t)} \sin \frac{\tB(t)}{2} \ket{S}.
\end{align}
A direct comparison yields the following:
\begin{align}
\label{keisu0}
C_N(t)&=e^{i\phi_N(t)} \cos \frac{\tB(t)}{2}, \nonumber \\
C_S(t)&=e^{i\phi_S(t)} \sin \frac{\tB(t)}{2}.
\end{align}

Because the BS provides that $0 \le \tB \le \pi$, both $\cos \frac{\tB}{2}$ and $\sin \frac{\tB}{2}$ are positive.
Accordingly, we have the following relationship between the zenith angle $\tB$ and expansion coefficients $C_N$ and $C_S$:
\begin{align}
|C_N|&= \cos \frac{\tB}{2}, \nonumber \\
|C_S|&= \sin \frac{\tB}{2}.
\end{align}
Therefore, we have,
\begin{align}
\tan \frac{\tB}{2} =\frac{\sin \frac{\tB}{2}}{\cos \frac{\tB}{2}}=\frac{|C_S|}{|C_N|}.
\end{align}
Then, the zenith angle is determined as follows:
\begin{align}
\tB=2 \tan^{-1} \left(\frac{|C_S|}{|C_N|}\right).
\end{align}
We obtain the phases $\phi_N$ and $\phi_S$ by 
\begin{align}
\phi_N &= \mathrm{Arg} (C_N), \nonumber \\
\phi_S &= \mathrm{Arg} (C_S).
\end{align}
Therefore, we obtain the azimuthal angle $\phiB$ in the BS representation, as follows:
\begin{align}
\phiB=\phi_S- \phi_N=\mathrm{Arg} (C_S)- \mathrm{Arg} (C_N) =\mathrm{Arg} \left(\frac{C_S}{C_N}\right).
\end{align}

\section{Application to the Rabi oscillation}

The TSRW approach provides analytical solutions for $C_N$ and $C_S$ using
\begin{align}
C_N&=\cos(|F|t) \cdot \mathrm e^{-i\omega_0 t}, \nonumber \\
C_S&=-i\frac{F}{|F|}\sin(|F|t) \cdot \mathrm e^{-i2\omega_0 t}.
\end{align}
Here, we explicitly express the dynamical term caused by the individual eigen frequency. Accordingly, a comparison with the BS representation provides us
\begin{align}
\label{CnCsharm}
C_N&=\mathrm e^{i\lambdaB} \cdot \cos \frac{\theta_\mathrm B}{2} =\cos(|F|t) \cdot \mathrm e^{-i\omega_0 t}, \nonumber \\
C_S&= \mathrm e^{i\lambdaB} \mathrm e^{i\phiB}\sin\frac{\theta_\mathrm B}{2} =
-i\frac{F}{|F|}\sin (|F|t)\cdot \mathrm e^{-i2\omega_0 t}.
\end{align}
The absolute value $|C_N|$ provides the BS zenith angle $\tB$, as follows:
\begin{align}
&\cos \frac{\theta_\mathrm B}{2} = |\cos \frac{\theta_\mathrm B}{2}|
=|\cos(|F|t)|, \nonumber \\
\therefore \quad &\tB=2\cos^{-1} \left(|\cos (|F|t) |\right) \Rightarrow 2|F|t.
\end{align}
The ratio of $|C_N|$ to  $|C_S|$ in Eq. \eqref{CnCsharm} gives the BS azimuthal angle $\phiB$;
\begin{align}
\mathrm e^{i\phiB} 
= -i\displaystyle{\left(\frac{F}{|F|}\right)}\frac{\tan (|F|t)}{|\tan(|F|t)|}\cdot \mathrm e^{-i\omega_0 t}.
\end{align}
Eventually, we obtain the BS azimuthal angle $\phiB$ as follows:
\begin{align}
\phiB&= -\omega_0 t - \frac{\pi}{2} \mathrm{Sign}[\tan (|F|t)]+\mathrm{Arg}(F) + 2n\pi.
\end{align}
Furthermore, we obtain the global phase as
\begin{align}
\lambdaB &=-\omega_0 t + \mathrm{Arg} \left(\frac{\cos|F|t}{|\cos|F|t|}\right).
\end{align}

\end{document}